\documentclass[sigconf]{acmart}

\setcopyright{rightsretained} 
 
\begin{document}

\author{Yan-Martin Tamm}
\email{yytamm@sber.ru}
\affiliation{%
  \institution{Sber AI Lab}
  \streetaddress{Oruzheiniy pereulok 41}
  \city{Moscow}
  \country{Russia}
  \postcode{127006}
}

\author{Rinchin Damdinov}
\email{rgdamdinov@sber.ru}
\affiliation{%
  \institution{Sber AI Lab}
  \streetaddress{Oruzheiniy pereulok 41}
  \city{Moscow}
  \country{Russia}
  \postcode{127006}
}

\author{Alexey Vasilev}
\email{avavasilyev@sber.ru}

\affiliation{%
  \institution{Sber AI Lab}
  \streetaddress{Oruzheiniy pereulok 41}
  \city{Moscow}
  \country{Russia}
  \postcode{127006}
}

\title{Quality Metrics in Recommender Systems: Do We Calculate Metrics Consistently?}

\begin{abstract}
Offline evaluation is a popular approach to determine the best algorithm in terms of the chosen quality metric. However, if the chosen metric calculates something unexpected, this miscommunication can lead to poor decisions and wrong conclusions. In this paper, we thoroughly investigate quality metrics used for recommender systems evaluation. We look at the practical aspect of implementations found in modern RecSys libraries and at the theoretical aspect of definitions in academic papers. We find that Precision is the only metric universally understood among papers and libraries, while other metrics may have different interpretations. Metrics implemented in different libraries sometimes have the same name but measure different things, which leads to different results given the same input. When defining metrics in an academic paper, authors sometimes omit explicit formulations or give references that do not contain explanations either. In 47\% of cases, we cannot easily know how the metric is defined because the definition is not clear or absent. These findings highlight yet another difficulty in recommender system evaluation and call for a more detailed description of evaluation protocols.
\end{abstract}

\copyrightyear{2021}
\acmYear{2021}

\acmConference[RecSys '21]{Fifteenth ACM Conference on Recommender Systems}{September 27-October 1, 2021}{Amsterdam, Netherlands}
\acmBooktitle{Fifteenth ACM Conference on Recommender Systems (RecSys '21), September 27-October 1, 2021, Amsterdam, Netherlands}\acmDOI{10.1145/3460231.3478848}
\acmISBN{978-1-4503-8458-2/21/09}



\maketitle

\section{Introduction}

Optimization of the right set of business performance metrics, such as click-through rate, time spent, items purchased or consumed, and the overall user engagement, is one of the critical tasks of a recommender system (RS).
The best way to determine whether an RS improves business metrics is to conduct an A/B test using live user feedback. However, an A/B test is usually a costly and time-consuming endeavor. Therefore, before conducting an A/B test, it is desirable to apply a cheaper and faster offline testing method to ensure no significant drop in the quality of provided recommendations. This lead to the development of different approaches such as testing on simulated data \cite{tian2020estimating}, off-policy evaluation \cite{gilotte2018offline, kato2020practical}, and offline evaluation \cite{carraro2020debiased}, which is a standard way to compare the performance of different models \cite{sun2020are}. It is crucial to maintain consistency and uniformity at all stages, including metric calculation. Inconsistent formulation of the quality metrics is a dangerous situation that can lead to improper model comparison and misinterpretation of the reported results. This problem is even more acute in the light of recent articles indicating concerns regarding the reproducibility and credibility of the reported results. 
In particular, some authors state that many results published at recent conferences on recommender systems are not easily reproducible, and those that are, usually can be outperformed by carefully tuned baselines. \cite{dacrema2019we, dacrema2021troubling, rendle2019difficulty}. Recommender systems is a challenging area for evaluation that has no universally accepted protocol, and a variety of methods exist, which means that different approaches are not directly comparable with each other \cite{meng2020exploring, canamares2020target, krichene2020sampled, sun2020are, canamares2020offline}.

This paper focuses on a careful examination of definitions and computations of various RecSys quality metrics introduced both in the literature and in the publicly available libraries. 
We demonstrate that metrics with the same name can be computed differently across libraries and papers, which is disturbing since it creates inconsistencies across the published results and claims in both academia and the industry. As a first step of addressing these inconsistencies, we systematize and summarize different approaches to calculating popular quality metrics. As a part of this process, we study how metrics are defined across various papers and whether or not they have precise definitions or refer to another publication. We conclude that 47\% of the studied papers do not have proper definitions, which is a disturbing number, requiring systematic re-examination of the RecSys performance evaluation process.

\section{Related work}

Implicit choices made during evaluation do affect results \cite{canamares2020offline} and should be brought to light with a comprehensive description of the protocol used. Variables at hand include dataset selection, data filtration, splitting strategy, choice of items to rank, metrics, and depth cut-offs. This observation also makes one wonder which option should be preferred over the other, and some recent papers try to answer this question.

Datasets used vary greatly from paper to paper \cite{dacrema2021troubling, sun2020are, meng2020exploring} and there is no justification for which dataset should be used that we know of, but it was shown that bigger versions of MovieLens are more robust to different filtering strategies \cite{tousch2019robust} and thus bigger datasets are preferable.

Different splitting strategies affect model comparison, and some authors argue that one should use global timestamp splitting to achieve a more fair and "closer to life" situation \cite{meng2020exploring, ji2021critical}.

Even metric calculation has no agreement among researchers as some calculate results using all the items available, some rank only items present in the test set, and some use sampling strategy to mix positive test set items with randomly sampled negative items. Some authors show that the usage of sampled metrics can lead to false conclusions if not carefully treated \cite{krichene2020sampled, li2020sampling} while others argue that some amount of sampling correlates more with a situation that would occur if we had missing at random data (MAR) \cite{canamares2020target}.

As we can see, the spectrum of different aspects of the evaluation is vast. In the following sections, we will look at different ways to interpret popular quality metrics and describe different variations found in libraries and papers.

\section{Evaluation Methodology}
\label{sec:eval}

In this study we use popular evaluation metrics such as HitRate, Precision, Recall, MRR, MAP, NDCG, and RocAuc. All the metrics, except for RocAuc, are calculated at depth cut-off 20. RocAuc was calculated using full predictions, except for the training data. Evaluation tools we used are our own library RePlay \cite{replay}, libraries released as a follow-up to the recently published papers discussing difficulties in RecSys evaluation, such as Beta RecSys \cite{meng2020beta}, DaisyRec \cite{sun2020are}, RecBole \cite{zhao2020recbole}, Elliot \cite{anelli2021elliot}, OpenRec \cite{yang2018OpenRec}, DL RS Evaluation \cite{dacrema2019we} and Python packages available at GitHub, such as MS Recommenders \cite{microsoft}, NeuRec \cite{wubinzzu}, RecSys PyTorch \cite{yoongi}, and rs\_metrics \cite{rsmetrics}.

To calculate metrics, we used the MovieLens-20m dataset and EASE recommender \cite{steck2019embarrassingly} to get predictions, as it is an easy model that produces a closed-form solution for a given regularization term. Ratings less than 4.5 were treated as negative and were removed. The global timestamp split was used so that 20\% of the latest ratings were in the test set. Users and items that appear only in the test set were removed. It should be noted that the choice of this particular model and the data is entirely arbitrary and could be treated as a black box. All that matters is that both the prediction and the test sets are fixed.

\section{Evaluation}
Results are reported in the table \ref{table:1}. Note that Precision and Recall are the only metrics with no disagreement in the resulting values across the libraries. In contrast, there are discrepancies in the MAP, MRR, NDCG, RocAuc, and HitRate metrics.

\begin{table*}[h!]
\captionsetup{justification=centering}
\centering
\begin{tabular}{ |c||c|c|c|c|c|c|c|  }
 \hline
 Library & HitRate@20 & MAP@20 & MRR@20 & NDCG@20 & Precision@20 & Recall@20 & RocAuc \\
 \hline
RePlay & 0.475 & 0.039 & 0.186 & 0.093 & 0.058 & 0.096 & 0.283 \\
DL RS Evaluation & 1.15 & 0.039 & 0.186 & 0.08 & 0.058 & 0.096 & 0.283 \\
DaisyRec & 0.475 & 0.023* & 0.275 & 0.093 & 0.058 & 0.096 & fail \\
Beta RecSys & — & 0.032 & — & 0.093 & 0.058 & 0.096 & 0.687 \\
RecBole & 0.475 & 0.039 & 0.186 & 0.093 & 0.058 & 0.096 & 0.687 \\
Elliot & 0.475 & 0.073 & 0.186 & 0.09 & 0.058 & 0.096 & 0.688 \\
OpenRec & — & — & — & 0.463 & 0.058 & 0.096 & 0.705 \\
MS Recommenders & — & 0.032 & — & 0.093 & 0.058 & 0.096 & 0.687 \\
NeuRec & 0.475* & 0.003 & 0.186 & 0.093 & 0.058 & 0.096 & — \\
rs\_metrics & 0.475 & 0.032 & 0.186 & 0.093 & 0.058 & 0.096 & — \\
 \hline
\end{tabular}

\caption{Metrics calculated using different libraries. \protect\\
* The calculation resulted in an error. If the required fix is minimal, we provide results with this mark.}
\label{table:1}
\end{table*}

The specifics of metric calculations will be discussed in the rest of this section. We will use the following notations:

\begin{itemize}
    \item $u$ is a user identificator
    \item $i$ is an item identificator
    \item $rec_k(u)$ is a recommendation list for user $u$ containing  top-k recommended items
    \item $rel(u)$ is a list of relevant items for user $u$ from the test set
    \item $rank(u, i)$ is a position of item $i$ in recommendation list $rec_k(u)$
    \item $ \mathbb{I}[\cdot] $ is an indicator function
\end{itemize}

If not stated otherwise, metrics are defined for a single user, and then they are averaged across all the users.

\subsection{Precision and Recall}
These are the only two metrics calculated the same in all of the libraries we tested, so we will just give formulas for reference.

\begin{displaymath}
Precision@k(u) = \frac{|rel(u) \cap rec_k(u)|}{k}
\end{displaymath}
\begin{displaymath}
Recall@k(u) = \frac{|rel(u) \cap rec_k(u)|}{|rel(u)|}
\end{displaymath}

\subsection{HitRate}

HitRate is defined as 1 if at least one item in the recommendation list was relevant and 0 otherwise. It can be expressed with the following formula:

\begin{displaymath}
HitRate@k(u) =\mathbb{I}[|rel(u) \cap rec_k(u)| > 0]
\end{displaymath}

Having a HitRate value being more than one in DL RS Evaluation in Table~\ref{table:1} is surprising. It turns out that the implemented HitRate is not the percentage of users for which we had at least one accurate prediction but the average number of correctly predicted items. This value also happens to be precisely $k=20$ times bigger than Precision. 

In this particular case, there are three different interpretations of this metric in one work. The implementation we just discussed, the definition in the paper \cite{dacrema2021troubling} and a reference for another paper \cite{zhang2018coupledcf}, provided for more information. The initial paper states that "The Hit Rate measures if a single hidden true positive was part of the top-n recommendation list", but the paper that is referenced to give the formal definition of HitRate proposes yet another definition for this metric, stating "$HR@K$: a recall-based measure, i.e.,\(HR@K = \frac{\#hits@K}{|GT|}\), to indicate whether the test item is in the top-K recommended item list ... where GT denotes the test list set" which happens to be the same thing as Recall.

\subsection{MRR}

MRR or the Mean Reciprocal Rank is the mean of the inverse position of the first relevant item. If no items were predicted correctly, it is defined as 0.
\begin{displaymath}
MRR@k(u)=\frac{1}{\min \limits_{i \in rel(u) \cap rec_k(u)} rank(u, i) }
\end{displaymath}

In DaisyRec, MRR calculates not the \emph{first} occurrence of the relevant item, but \emph{sum of all} inverse positions of relevant items. This means that the value of the metric can be bigger than 1. In the case of leave-one-out splitting, it does not matter, and the results will be identical with other libraries.

\subsection{MAP}

Mean Average Precision is averaged across users, and all the differences occur in the calculation of the Average Precision. 

\begin{displaymath}
AP@k(u) = \frac{1}{x} \sum_{i \in rec_k(u)} \mathbb{I}[i \in rel(u)]Precision@rank(u,i)(u)
\end{displaymath}

Where $x$ is the averaging term that can be one of the following:

\begin{itemize}
    \item $k$, the number of recommendations
    \item $r=|rel(u)|$, the total number of relevant items
    \item $min(k, r)$
\end{itemize}

There are five different MAP values, which is the biggest spread among all the metrics we tested. The initial version of AP was meant to represent the area under the precision-recall curve. In this version $x=r$, the number of relevant items for user $u$. This gives us a value of 0.032 as in Beta Recsys, MS Recommenders, and rs\_metrics. 

DaisyRec uses $x=k$, the length of the recommendation list, which in our case is 20. This is how we get 0.023. 

If the number of relevant items is greater than $k$, no possible ranking will achieve metric value 1. Furthermore, if $r < k$, but the formula uses $x=k$, then the metric will punish and lower the results even they cannot be improved under this test set. This is the motivation to use $x=min(k,r)$. This is the variant implemented in RecBole, RePlay, and DL RS Evaluation, which yields 0.039. Elliot uses $x=k$, but in this case the term $\mathbb{I}[i \in rel(u)]$ is dropped completely, so the summation is over \emph{all} $Precision@k$ terms which is strange, because documentation states to use $x=min(k,r)$ and have $\mathbb{I}[i \in rel(u)]$ present. This contributes to a value of 0.073. NeuRec intends to use $x=min(k,r)$, but a bug changes this term. The metric is calculated for all cut-offs simultaneously and constructs a list for $x$ for every possible cut-off up to $k$. The bug is in the applied cumulative sum function over $x$ that corrupts the intended value. This is why the result is much lesser than all the other variants with just 0.003.

\subsection{NDCG}
Normalized Discounted Cumulative Gain is a ranking metric that can accept arbitrary relevance values $rating(u, i)$ to weigh the results. In RecSys, item ratings from the test set are usually used for this purpose.

\begin{displaymath}
NDCG@k(u) = \frac{DCG@k(u)} {IDCG@k(u)}
\end{displaymath}
\begin{displaymath}
DCG@k(u) = \sum_{i \in rec_k(u)} \frac{2^{rating(u, i)}-1} {\log_2(rank(u, i)+1)}
\end{displaymath}

where $IDCG@k$ is the biggest possible value of $DCG@k$, that is the value we get if we treat the sorted test set as a prediction of length $k$. We will call this version the Weighted NDCG (WNDCG).

In RecSys relevance is usually binary, 1 if the item is relevant and 0 otherwise, in this case the formula can be rewritten as 
\begin{displaymath}
DCG@k(u) =  \sum_{i \in rec_k(u)} \frac{\mathbb{I}[i \in rel(u)]} {\log_2(rank(u, i)+1)}
\end{displaymath}

We will call this version the Binary NDCG (BNDCG).

All the libraries can be divided into two groups: those that calculate BNDCG and those that calculate WNDCG. Beta RecSys, RecBole, RePlay, DaisyRec, MS Recommenders, NeuRec, and rs\_metrics completely ignore relevance values and treat them as 1. The value for BNDCG is 0.093. As for OpenRec, 0.463 comes from a bug that the normalization term is missing, so the resulting metric is not BNDCG but just BDCG.

DL RS Evaluation and Elliot support arbitrary relevance values and calculate WNDCG. In our case, there are two possible values for the rating in the test set: 4.5 and 5.0. Elliot has 0.09, which is a correct value for the WNDCG formula that uses 2 to the power of the rating. DL RS Evaluation has 0.08 because it uses natural logarithm instead of logarithm with base 2.

\subsection{RocAuc}

RocAuc or Reciever Operating Characteristic Area Under Curve \cite{bradley1997use} is the most interesting case because it comes from a classification background, and it is not completely clear how to adapt it for a recommendation scenario where the output of a model is not a binary class, but a list of ranked item ids. In classification, a classifier that outputs labels of 1 and 0 at random would get a RocAuc score of 0.5, but a RecSys model that outputs random item ids as its prediction will probably not reach 0.5. The intuition behind RocAuc as a comparison with random predictions does not apply in RecSys as the model can easily get a score below 0.5. Two main decisions are
\begin{itemize}
    \item do we use full ranking for all items or just truncated top-k lists?
    \item do we calculate a metric for all predictions at once or do we average AUCs across users as we usually do for other metrics?
\end{itemize}

Further information on which  variants are available in various libraries is provided in Table~\ref{table:2}

\begin{table}[h!]
\captionsetup{justification=justified}
\centering
\begin{tabular}{ |c||c|c|c|c|  }
\hline
 Library & SAUC & GAUC & GAUC@20 & LAUC@20 \\
 \hline
RePlay & — & — & 0.283 & — \\
DL RS Evaluation & — & — & 0.283 & —  \\
DaisyRec & — & — & fail & — \\
Beta RecSys & 0.687 & — & — & — \\
RecBole & 0.687 & 0.858 & — & — \\
Elliot & 0.688 & 0.705 & — & 0.112 \\
OpenRec & — & 0.705 & — & — \\
MS Recommenders & 0.687 & — & — & — \\
\hline
\end{tabular}

\caption{AUC values distributed by the meaning of the calculated metric.}
\label{table:2}
\end{table}

The classical way is to use full rankings and stack all the predictions together as if it was a classification problem. We will call this variant the Stacked AUC (SAUC). It is implemented in Beta RecSys, RecBole, and MS Recommenders; this variant has a value of 0.687. Although Elliot intends to use SAUC, in reality, it introduces another variant since it does not stack \emph{predictions} but instead calculates the area of rectangles contributing to the AUC for each user separately and then averages these intermediate terms, resulting in 0.688.

This raises the question of why we should blend recommendations for different users at all. Users do not know anything about other users' recommendations. If we choose to average AUCs calculated for each user, we get a more natural variant of the Group AUC (GAUC) metric implemented in OpenRec, DL RS Evaluation, Elliot, RePlay, and RecBole. In OpenRec and DL RS Evaluation, it uses the name AUC. OpenRec and Elliot accept full rankings and thus achieve 0.705. RecBole has GAUC 0.858 because it uses the number of rated items as a weight to this user's AUC score and skips users that only have either all items predicted correctly or no correctly predicted items at all. Other libraries assign 0 in this situation. DL RS Evaluation and RePlay apply cut-off in the evaluator before passing data further into the metric, which gives us 0.283.

Some libraries implement more than one variant of RocAuc. For example, Elliot supports the AUC, GAUC, and LAUC metrics \cite{schroder2011setting}. LAUC uses a top-k recommendations list instead of full rankings but takes into account all the items. The motivation is that we usually do not care as much about rankings at the end of a list as we do at the beginning.

\section{Paper Analysis}

We have shown that there is room for misinterpretation of metrics in their implementations, but what is the situation regarding academic papers? Is there a variation in descriptions, and can a reader know what introduced metrics are supposed to mean? To this end, we collected papers introducing new models, some of them being long-known baselines, some stating to be state-of-the-art \cite{hu2018leveraging, he2017neural, yu2014personalized, he2018outer, he2015trirank, cheng2018delf, ebesu2018collaborative, liang2018variational, li2017collaborative, xue2017deep, wang2015collaborative, zhang2018coupledcf, zheng2018spectral, tay2018latent, ning2011slim}. 

Results are summarized in table \ref{table:3}. It turns out that only 33\% of papers (5/15) explicitly provide a full description of used metrics. 20\% (3/15) gives a proper reference to another article that has a sufficient description. That gives us 53\% (8/15) of papers for which we can be sure which versions of metrics authors intended to use, while in 47\% (7/15), we cannot be so sure, because there is only a partial description or a reference to a paper that has no formulas either. 



\begin{table}[h!]
\captionsetup{justification=justified}
\centering
\begin{tabular}{ |c|c|c|c|c|c|c|c|  }
 \hline
 Wrong ref. & Correct ref. & Partial desc. & Full desc.  \\
 \hline
4/15 & 3/15 & 3/15 & 5/15 \\
 \hline
\end{tabular}

\caption{This table shows how metrics are introduced in articles.}
\label{table:3}
\end{table}

Most of the wrong references are to NeuMF paper \cite{he2017neural} that gives only a brief description "the HR intuitively measures whether
the test item is present on the top-10 list, and the NDCG
accounts for the position of the hit by assigning higher scores
to hits at top ranks." NeuMF does reference TriRank paper \cite{he2015trirank}, which has exact formulas for HitRate and NDCG, but we find this to be inconvenient for the reader to follow more than one link in a chain to find the reference. It is also interesting that the version of HitRate introduced in the TriRank paper calculates the same thing that does Recall from the formulas we provided earlier. It also contradicts the meaning that is provided in the text of the NeuMF paper. This was not counted as the wrong reference because there was a formula present, but a reader may only wonder which variant is used until he checks the source code. CoupledCF \cite{zhang2018coupledcf} is another paper that introduces a "Recall" variant of HitRate.

Some papers introduce variations to the discussed metrics. For example, the SLIM paper \cite{ning2011slim} has the formula for the Average Reciprocal Hit Rate (ARHR), which is a synonym for MRR we introduced earlier. This name can be found in earlier works too \cite{deshpande2004item}, but it seems not to be widely used today. Mult-VAE \cite{liang2018variational} introduces the formula for Recall that was not present in any other paper or library we have discussed. Following the same logic as with the averaging term in AP, the denominator in the Recall formula is replaced from the total number of relevant items $r$ to $min(r,k)$ so that metric can reach 1 when we calculate it using predictions of length $k$.

CDL \cite{wang2015collaborative} is an interesting case where the definition for a simple Recall metric is present, but when it comes to a fairly complex MAP, there is no explanation except for "Another evaluation metric is the mean average precision (mAP)". This is quite confusing because we have seen three different variants in the libraries.

\section{Conclusion}

This paper examined various RecSys metrics across different Python libraries and identified several variations of these metrics; supplemental code is available on GitHub \cite{repo}. We showed that there is no consensus on the exact meaning of popular quality metrics and that the more complex a metric is, the more room there is for different interpretations of the metric. This provides an additional incentive to describe the evaluation protocol \cite{canamares2020offline} more thoroughly and comprehensively. The key takeaways that we found are:

\begin{itemize}
    \item Precision is one metric everyone agrees upon
    \item sometimes metric variants are not bounded by 1, as with HitRate in the DL RS Evaluation and MRR in DaisyRec
    \item there is uncertainty on how to average metrics when the number of relevant items $r$ is not equal to the length of the recommendation list $k$ for some users. If $r > k$, then Recall won't reach 1 for any recommendation list of length $k$. Moreover, if $r < k$, then AP will be too pessimistic if averaged by $k$ in this situation.
    \item there are two variations for NDCG, the weighted and the binary, BNDCG being more popular. Sometimes the logarithm in the formula has a different base from 2.
    \item AUC is the most disputed metric. Whether to calculate it using all ranked items, or just the top $k$, stack all predictions to calculate AUC, average by users with GAUC, or use LAUC.
\end{itemize}

While some metric variants are popular and almost universally accepted, such as NDCG using binary relevance, or HitRate and MRR being limited by 1, others are not so straightforward. 
Therefore, it is crucial for the RecSys community to standardize definitions of various metrics or explicitly recognize and accept their variants. 

\begin{acks}
We would like to express our gratitude to Alexander Tuzhilin for text revision and stylistic edits, and to Gleb Gusev, Bulat Ibragimov, and Anna Volodkevich for valuable discussions and comments.
\end{acks}

\bibliographystyle{ACM-Reference-Format}
\bibliography{reference}

\end{document}